\documentclass[aps,pra,twocolumn,superscriptaddress,showpacs,10pt]{revtex4-1}
\usepackage{amsmath,amsfonts,amssymb,bbm,graphicx}

\newcommand{\eqn}[1]{\begin{equation} #1 \end{equation}} % equation environment
\newcommand{\aln}[1]{\begin{align} #1 \end{align}}       % align environment
\newcommand{\mc}{\mathcal}                               % script letters
\newcommand{\mbf}{\mathbf}                               % bold letters
\newcommand{\eq}[1]{(\ref{#1})}              % equation reference
\newcommand{\pd}{\partial}                   % partial derivative sign
\newcommand{\wtilde}{\widetilde}             % wide tilde
\renewcommand{\l}{\left}
\renewcommand{\r}{\right}
\newcommand{\Tr}{\text{Tr}\,}                % Tr
\newcommand{\q}{\text q}                     % some notation
\newcommand{\cl}{\text {cl}}
\newcommand{\A}{\text A}
\newcommand{\R}{\text R}
\newcommand{\K}{\text K}
\newcommand{\Q}{\text Q}

\begin{document}

% Use the \preprint command to place your local institutional report
% number in the upper righthand corner of the title page in preprint mode.
% Multiple \preprint commands are allowed.
% Use the 'preprintnumbers' class option to override journal defaults
% to display numbers if necessary
%\preprint{}

%Title of paper
\title{Nonlinear sigma model for optical media with linear absorption or gain}

% repeat the \author .. \affiliation  etc. as needed
% \email, \thanks, \homepage, \altaffiliation all apply to the current
% author. Explanatory text should go in the []'s, actual e-mail
% address or url should go in the {}'s for \email and \homepage.
% Please use the appropriate macro foreach each type of information

% \affiliation command applies to all authors since the last
% \affiliation command. The \affiliation command should follow the
% other information
% \affiliation can be followed by \email, \homepage, \thanks as well.
\author{Zhong Yuan Lai}
%\email[]{Your e-mail address}
%\homepage[]{Your web page}
%\thanks{}
%\altaffiliation{}
\affiliation{Physikalisches Institut, Universit\"at Bonn, Nu{\ss}allee 12, 53115 Bonn, Germany}

\author{Oleg Zaitsev}
\affiliation{Department of Information Display, Kyung Hee University,
1 Hoegi-dong, Dongdaemun-gu, Seoul 130-701, Korea}
\thanks{Corresponding author}
\email[]{ozaitsev@khu.ac.kr}

%Collaboration name if desired (requires use of superscriptaddress
%option in \documentclass). \noaffiliation is required (may also be
%used with the \author command).
%\collaboration can be followed by \email, \homepage, \thanks as well.
%\collaboration{}
%\noaffiliation

%\date{\today}

\begin{abstract}

In the framework of the Keldysh technique, we formulate the nonlinear sigma model for disordered optical media with linear absorption or gain. The effective action for fluctuations of the matrix field about the saddle point acquires an extra term due to the nonconservative nature of the system. We determine the disorder-averaged Green-function correlator, which has a diffusion pole modified by a finite absorption/gain rate. The diffusion coefficient is found to be close to its value for conservative systems in the relevant range of parameters. In the medium with gain, the random-lasing threshold depends on the sample size.

\end{abstract}

% insert suggested PACS numbers in braces on next line
% insert suggested keywords - APS authors don't need to do this
%\keywords{}

%\maketitle must follow title, authors, abstract, \pacs, and \keywords
\maketitle

% body of paper here - Use proper section commands
% References should be done using the ~\cite, \ref, and \label commands

\section{Introduction}

The transport of waves through disordered matter has been a topic of recurring interest ever since the discovery of the Anderson localization in electronic systems~\cite{ande58}. Analogous phenomena have been subsequently studied for the transport of classical~\cite{john83,ande85,mark88,kroh93}, matter~\cite{roat08,bill08}, and even seismic waves~\cite{laro04}. 

The research on classical-wave propagation in disordered media has been motivated by the conjectured possibility of the localization of light. The results, such as the enhancement of dwell times due to resonant scatterers and, hence, lower energy-transport velocities~\cite{lage96} and the correction term in the Ward identity due to frequency-dependent scattering potentials~\cite{bara91}, have shown that, while retaining many similarities, the behavior of light in disordered media differs from that of electrons in several important aspects. One of these aspects concerns the propagation of light in nonconservative disordered media. Such systems can be physically realized, for example, as random lasers~\cite{cao05,wier08,zait10}, which have received much attention recently. A promising research direction in this context are theories that combine description of wave propagation through disordered medium with the nonlinear laser equations~\cite{flor04,fran09,fran09b}.

The properties of light diffusion in absorbing media was studied using the photon transport equation~\cite{durd97,furu94,furu97}. In particular, it was argued that, in the parameter range of validity of the diffusion equation, the diffusion coefficient is close to its value in the conservative medium. The treatment of light propagation starting from the wave equation has been mainly conducted via the self-consistent diagrammatic theory~\cite{voll80b,voll92}. Interesting results, such as corrections to the bare diffusion coefficient due to the additional terms in the Ward identity~\cite{luba05,fran06} and dynamics of Anderson localization in quasi-one-dimensional geometry~\cite{skip04} and open three-dimensional media~\cite{skip06} have been obtained by these methods. An alternative description of classical wave propagation is provided by the so called effective models of disordered systems, commonly known as the nonlinear sigma model~\cite{wegn79,scha80} (NLSM). Being originally developed for electronic systems, the (supersymmetric) NLSM describing light propagation in a \emph{conservative} disordered medium was derived in Refs.~\cite{john83,elat98}. Later, the effects of an open boundary on the diffusion coefficient were studied~\cite{tian08} by using a similar model. Unlike the self-consistent theory of transport, effective models have not yet been applied to optical systems with absorption or gain. The effective models can be useful, e.g., in describing special properties of light localization in such systems~\cite{deyc01,payn10,paas96}.

In the present work we formulate the Keldysh nonlinear sigma model~\cite{kame09} for the propagation of electromagnetic waves in \emph{nonconservative} disordered media in the diffusive regime. Systems with absorption or gain are relatively simple to treat in the Keldysh formalism, which makes it possible to define an action needed for the field-theoretical description. By following the general scheme as outlined in Ref.~\cite{kame09} for electronic systems, we derive an effective NLSM action where we obtain a term due to nonconservativeness of the medium. A similar contribution was found in Ref.~\cite{tian08}; in that case the term originated from the openness of the system. 

Furthermore, by using the standard methods~\cite{kame09}, we show that the light propagation can be described by a diffusion equation for nonconservative medium. The conditions under which the NLSM yields the diffusion equation are found to be equivalent to the restrictions imposed in the theory of transport equation~\cite{durd97}. Similarly, the diffusion coefficient that we derive is almost independent of the absorption or gain under these conditions. For the amplifying medium, we discuss the applicability of the linear-gain approximation and determine the threshold of random lasing.

\section{Keldysh approach to light propagation}

\subsection{Partition function for nonconservative medium}

We consider an optical medium defined by a complex dielectric constant $\epsilon (\mbf r, \omega)$. Restricting our theory to the TM~modes in two dimensions, we describe the electric field by its normal component $E_\omega (\mbf r) = i \omega A_\omega (\mbf r)$ (in the Coulomb gauge), where $A_\omega (\mbf r)$ is the normal component of the vector potential. We use the Gaussian units with the velocity of light $c \equiv 1$. The transversality condition $\mbf \nabla \cdot (\epsilon \mbf A) = 0$ with $\epsilon$ varying in two dimensions, leads to the two-dimensional wave equation
\eqn{
  [\mbf \nabla^2 + \epsilon (\mbf r, \omega)\, \omega^2] A_\omega (\mbf r) 
  = 0.
\label{wave_eq}
}

For real $\epsilon (\mbf r, \omega)$, this equation, and its complex conjugate, can be obtained by setting to zero the functional derivatives 
\eqn{
  \frac {\delta S} {\delta A} = 0, \quad \frac {\delta S} {\delta A^*} = 0
}
of the action (Hamilton principal function)
\aln{
  &S[A,A^*] \notag \\
  &= \frac 1 {16 \pi} \int\! d \mbf r\, \frac {d \omega} {2 \pi}
  \l[\epsilon (\mbf r, \omega)\, \omega^2 |A_\omega (\mbf r)|^2 -
  |\mbf \nabla A_\omega (\mbf r)|^2 \r] \notag \\
  &= \frac 1 {16 \pi} \int\! d \mbf r\, \frac {d \omega} {2 \pi} 
  A^*_\omega (\mbf r) \l[\epsilon (\mbf r, \omega)\, \omega^2 + \mbf \nabla^2
  \r] A_\omega (\mbf r),
}
treating $A$ and $A^*$ as independent functions. The action can be rewritten in the representation-free operator notation~as
\eqn{
  S [A, A^\dag] = \frac 1 {16 \pi} A^\dag G^{-1} A,
}
where the inverse Green function operator $G^{-1} = \epsilon (\mbf r, \omega)\, \omega^2 + \mbf \nabla^2$ in the $(\mbf r, \omega)$ representation and $A$ ($A^\dag$) is the Hilbert-space vector $A_\omega (\mbf r)$ [$A_\omega^* (\mbf r)$].

In order to construct the quantum Hamiltonian, one expresses the energy of the system in terms of the vector potential. $A_\omega (\mbf r)$ and $A_\omega^* (\mbf r)$ are then expanded in the normal modes of the system, the expansion coefficients become the photon annihilation and creation operators. 

In the Keldysh field-theoretical approach~\cite{kame09,altl10} we calculate the partition function
\eqn{
  Z = \Tr (U \rho),
}
where $\rho$ is the density matrix at time $t = - \infty$, with $\Tr \rho = 1$, and 
\eqn{
  U = \text T_{\mc C} \exp \l[ - i \int_{\mc C} H (t)\, dt \r], \quad 
  \hbar \equiv 1,
}
is the time-evolution operator along the Keldysh contour~$\mc C$. The contour begins at $t = - \infty$, where the state of the system is known, then goes forward in time up to $t = \infty$, where it turns back and goes to $t = - \infty$. $\text T_{\mc C}$~denotes the time ordering along the contour. The Hamiltonian $H (t)$ is switched on adiabatically, starting from a trivial Hamiltonian~$H (- \infty)$.  $H (t)$~is the same on the forward and backward branches of the contour. This condition leads to $U = 1$, and, hence, $Z = 1$. 
If the source terms that are different on the two branches are added to the Hamiltonian then $Z \ne 1$. The (functional) derivatives of the type $\delta Z[J] / \delta J |_{J=0}$ with respect to the sources~$J$ generate averages with the density matrix propagated from $t = - \infty$ to relevant times.

The partition function can be written in the form of a functional integral over the fields (i.e., the classical functions) $A$ and~$A^\dag$. To this end, we represent the classical action along the Keldysh contour~as
\aln{
  S_{\mc C} &= \frac 1 {16 \pi} [ A^\dag_+\, G^{-1} A_+ - A^\dag_-\, 
  G^{-1} A_- ] \notag \\
  &= \frac 1 {16 \pi} [(A^\cl)^\dag\, G^{-1} A^\q + (A^\q)^\dag\, 
  G^{-1} A^\cl],
\label{Sc}
}
where the subscripts ``$\pm$'' denote the fields on the forward and backward branches of the contour and the so called classical and quantum fields are defined by
\eqn{
  A^\cl = \frac 1 {\sqrt 2} (A_+ + A_-), \quad 
  A^\q = \frac 1 {\sqrt 2} (A_+ - A_-).
\label{rot}
}
The minus sign in front of the $A^\dag_-\,  G^{-1} A_-$ term in Eq.~\eq{Sc} takes care of the time reversal on the backward branch, whereas $A_-$ is the representation-free (vector) notation for the function $A_- (\mbf r, t)$ with the forward time ordering. It is convenient to consider $A^\cl$ and $A^\q$ as components of a single field 
\eqn{
  \hat A = \l(\begin{array}{cc} A^\cl \\ A^\q \end{array} \r)
}
in the Keldysh space, which is twice the size of the original Hilbert space. (We will furnish the vectors and operators in this space with a hat.) Then the contour action can be written in the form (dropping the subscript~``$\mc C$'')
\eqn{
  S [\hat A, \hat A^\dag] = \frac 1 {16 \pi} \hat A^\dag\, \hat G^{-1} \hat A,
}
where $\hat G^{-1}$ has a $2 \times 2$ matrix structure with zeros on the diagonal and equal off-diagonal elements. 

In order to use $S [\hat A, \hat A^\dag]$ in the functional integral for~$Z$, the operator $\hat G^{-1}$ has to be regularized~\cite{kame09} by imposing the causality structure on its matrix:
\aln{
  &\hat G^{-1} = \l(\begin{array}{cc} 0 & (G^{-1})^\A \\ (G^{-1})^\R &
  (G^{-1})^\K \end{array} \r), 
\label{G-1_matr}\\
  &(G^{-1})^{\R,\A}_\omega (\mbf r) = \epsilon (\mbf r, \omega)\, \omega^2 
  + \mbf \nabla^2 \pm i0^+, 
\label{G-1_real_e}\\
  &(G^{-1})^\K = (G^{-1})^\R F - F (G^{-1})^\A.
\label{G-1_K}
}
Here, $(G^{-1})^{\R,\A,\K}$ are the retarded, advanced, and Keldysh components of the inverse Green function operator. The operator $F$ that parameterizes $(G^{-1})^\K$ depends on the thermal distribution. Equation~\eq{G-1_real_e} is written under assumption of real~$\epsilon (\mbf r, \omega)$. In the medium with absorption, it is generalized~to
\aln{
  &(G^{-1})^{\R,\A}_\omega (\mbf r) = \epsilon' (\mbf r, \omega)\, \omega^2 
  + \mbf \nabla^2 \pm i \epsilon'' (\mbf r, \omega)\, \omega^2, 
\label{G-1_RA}\\
  &\epsilon' (\mbf r, \omega) = \text{Re}\, [\epsilon (\mbf r, \omega)], \quad
  \epsilon'' (\mbf r, \omega) = \text{Im}\, [\epsilon (\mbf r, \omega)],
}
where $\epsilon'' > 0$. In the case of the gain medium, $\epsilon'' < 0$, the time-dependent Green function is exponentially diverging, and its Fourier transform to the frequency domain does not exist. The frequency representation can be defined with the help of the Laplace transform, which is equivalent to introducing a fictitious absorption to the system. When the results of a calculation are transformed back to the time representation, they should not depend on the fictitious absorption. This means that one can perform the calculations in the frequency domain assuming $\epsilon'' > 0$, and obtain the final results by analytic continuation to $\epsilon'' < 0$.

The functional-integral representation of the partition function becomes 
\eqn{
  Z = \mc N \int\! D[\hat A, \hat A^\dag]\, e^{iS [\hat A, \hat A^\dag]},
\label{Z}
}
where $\mc N$ is the nonessential normalization constant that ensures $Z = 1$ and the measure is defined~by
\eqn{
  D[\hat A, \hat A^\dag] = \prod_{\substack{\mbf r, \omega,\\j = \cl, \q}}
  \frac {d \bigl(\text{Re}\, A^j_\omega (\mbf r) \bigr)\,
  d \bigl(\text{Im}\, A^j_\omega (\mbf r) \bigr)} \pi. 
\label{DA}
}
Here and below we set to unity the step size for the grid used to discretize the integral.

We note that the Keldysh formalism is especially appropriate for the description of systems with absorption or gain as it naturally takes into account the finite $\epsilon'' (\mbf r, \omega)$ in the causality structure of the inverse Green function, Eqs.~\eq{G-1_matr}, \eq{G-1_RA}, and~\eq{G-1_K}.

\subsection{Disorder average}

We shall study the effect of disorder in the refractive index and assume the absorption or gain in the system to be spatially uniform. Hence, we represent the dielectric constant in the form
\eqn{
  \epsilon (\mbf r, \omega) = \epsilon' (\omega) + \Delta 
  \epsilon' (\mbf r, \omega) + i \epsilon'' (\omega)
}
with the averages over disorder realizations \mbox{$\langle \Delta \epsilon' (\mbf r, \omega) \rangle = 0$} and $\langle \Delta \epsilon' (\mbf r, \omega)\, \Delta \epsilon' (\mbf r', \omega) \rangle \propto \delta (\mbf r - \mbf r')$. 

In order to define the scattering time~$\tau$, let us, for a moment, neglect~$\epsilon'' (\omega)$. The wave equation~\eq{wave_eq} can be interpreted as a time-independent Schr{\"o}dinger equation with the energy $\mc E (\omega) = \epsilon' (\omega)\, \omega^2$ and the potential energy $V (\mbf r, \omega) = -\Delta \epsilon' (\mbf r, \omega)\, \omega^2$. The scattering time~$\tau$ and other characteristic time scales of the system are assumed to be much larger than~$\omega_0^{-1}$, where $\omega_0$ is the typical optical frequency. The slowly varying amplitude $\wtilde A (\mbf r, t) = A (\mbf r, t) \exp (i \omega_0 t)$ satisfy the approximate equation
\eqn{
  i \wtilde \hbar \frac {\pd \wtilde A} {\pd t} = [-\mbf \nabla^2 + 
  V (\mbf r, \omega_0) - \mc E (\omega_0)] \wtilde A,
}
which is the time-dependent Schr{\"o}dinger equation with the ``optical Planck constant''
\eqn{
  \wtilde \hbar = \frac {d \mc E (\omega_0)} {d \omega_0}.
}
The scattering time can now be defined by analogy with the quantum scattering time via the correlation function~\cite{akke07}
\eqn{
  \langle V (\mbf r, \omega_0)\, V (\mbf r', \omega_0) \rangle = 
  \frac {\wtilde \hbar} {2 \pi \nu \tau}\, \delta (\mbf r - \mbf r'),
}
where $\nu = dn/d \mc E$ is the quantum density of states (per unit volume). Note that $\nu = \nu_0/\wtilde \hbar$, where $\nu_0 = dn/d \omega$ is the standard optical density of modes. 

A disorder average of the partition function can be obtained by evaluating the functional integral
\aln{
  &\langle Z \rangle = \int D [V]\, Z\, \exp \l[ - \frac {\pi \nu \tau} 
  {\wtilde \hbar} \int d \mbf r\, V^2 (\mbf r, \omega_0) \r], \\
  &D [V] = \prod_{\mbf r} \sqrt {\frac {\nu \tau} {\wtilde \hbar}} \,
  d V (\mbf r, \omega_0).
}
The disorder-dependent part of the action~is
\aln{
  &\Delta S[\hat A, \hat A^\dag, V] = -\frac 1 {16 \pi} \hat A^\dag V 
  \hat \gamma \hat A, \quad
  \hat \gamma \equiv \l( \begin{array}{cc} 0 & 1 \\ 1 & 0 \end{array} \r),
}
where $V (\mbf r, \omega_0)$ appears as an operator~$V$ diagonal in~$\mbf r$; $F$~is assumed to be diagonal in~$\mbf r$, as well. By completing the square, we obtain the disorder contribution to the partition function
\eqn{
  \langle e^{i \Delta S} \rangle = \exp \l[ - \frac {\wtilde \hbar} 
  {4 \pi \nu \tau}\! \int\! d \mbf r \l(\frac 1 {16 \pi} 
  \hat A^\dag (\mbf r) \hat \gamma \hat A (\mbf r) \r)^2 \r].
\label{e_idS}
}
The short-hand notation $\hat A (\mbf r)$ is used for the Keldysh-space vector $\hat A$ with the fixed index~$\mbf r$, i.e., it is a vector in the space with the reduced dimensionality; in this notation, $\hat A^\dag (\mbf r) \hat A (\mbf r)$ involves a summation over the remaining indices, e.g., $\omega$ and the Keldysh index. 

The negative sign in the exponent~\eq{e_idS} is essential for the properties of nonlinear sigma model in the optical medium. In contrast to a fermionic system, the sign cannot be changed by commuting the fields.

\section{Nonlinear sigma model}

\subsection{Hubbard-Stratonovich transformation. Saddle point}

The term of the fourth-order in the fields in Eq.~\eq{e_idS} can be converted to a second-order term with the help of the Hubbard-Stratonovich transformation yielding
\aln{
  &\exp \l[ - \frac {\wtilde \hbar} {4 \pi \nu \tau}\! 
  \int\! d \mbf r \l(\frac 1 {16 \pi} 
  \hat A^\dag (\mbf r) \hat \gamma \hat A (\mbf r) \r)^2 \r] =\notag \\
  &\mc N_Q \int D [\hat Q] 
  \exp\! \l[ - \frac {\pi \nu \wtilde \hbar} 
  {4 \tau}  \Tr \hat Q^2 + i  \frac {\wtilde \hbar} {32 \pi \tau}
  \hat A^\dag \hat \gamma \hat Q \hat A \r], 
\label{HS}\\
  &\Tr \hat f \equiv \sum_{j=\cl,\q} \int d \mbf r\, \frac 
  {d \omega} {2 \pi}\, f^{jj}_{\omega \omega} (\mbf r).
}
The auxiliary field $\hat Q$ is the Hermitian operator diagonal in~$\mbf r$. The measure~$D [\hat Q]$ is defined over the independent matrix elements by analogy to Eq.~\eq{DA}. The normalization constant~$\mc N_Q$ is determined by setting $\hat A = 0$ and $\hat A^\dag = 0$. The negative coefficient in front of $\Tr \hat Q^2$ determines the scale of~$\hat Q$ and can be chosen freely. The present choice leads to the simple form of matrix~$\hat \Lambda$ introduced in Eq.~\eq{Lambda}. In order to prove Eq.~\eq{HS} more easily, one can define the matrix  $\hat {\mc A} (\mbf r) = \hat A (\mbf r) \otimes \hat A^\dag (\mbf r)$ where the tensor product applies to the Keldysh and $\omega$ subspaces. Then one represents
\eqn{
  \hat A^\dag \hat \gamma \hat Q \hat A = \Tr (Q\, \hat {\mc A}\, \hat \gamma).
}
Now the field $\hat Q$ can be integrated out after completing the square. 

Using Eqs.~\eq{e_idS} and~\eq{HS} in Eq.~\eq{Z} and integrating out the fields $\hat A$ and~$\hat A^\dag$, we obtain the disorder-averaged partition function
\aln{
  &\langle Z \rangle = \wtilde {\mc N}_Q \int D [\hat Q]\, e^{i S[\hat Q]}, 
\label{Zdis}\\
  &i S[\hat Q] \equiv - \Tr \l[ \frac {\pi \nu \wtilde \hbar} {4 \tau}  \hat
  Q^2 + \ln \l(\hat G_0^{-1} + \frac {\wtilde \hbar} {2  \tau} \hat
  \gamma \hat Q \r) \r],
\label{SQ}
}
where $\hat G_0^{-1}$ is the inverse Green function operator that does not include the disordered part of the dielectric constant and all $\hat Q$-independent factors are included in the normalization constant~$\wtilde {\mc N}_Q$. 

In the limit of large scattering time, the main contribution to $\langle Z \rangle$ comes from the neighborhood of a saddle point. The saddle-point equation 
\eqn{
  \hat Q (\mbf r) \hat \gamma = - \frac 1 {\pi \nu} \l(\hat G_0^{-1} + 
  \frac {\wtilde \hbar} {2  \tau} \hat \gamma \hat Q \r)^{-1}_{\mbf r \mbf r}
\label{saddle}
}
follows from the condition of stationary variation of $S[\hat Q]$ with respect to~$\hat \gamma \hat Q$. In the $(\mbf k, \omega)$ representation, 
\eqn{
  (G_0^{-1})^{\R,\A}_\omega (\mbf k) = \mc E (\omega) - k^2 \pm i
  \epsilon''(\omega)\, \omega^2
\label{G_0-1_RA}
}
is diagonal. We will look for the solutions $Q^{\R,\A}_\omega$ in the $(\cl, \cl)$ and $(\q, \q)$ blocks of~$\hat Q$, respectively, which are uniform in~$\mbf r$ and diagonal in~$\omega$. Equation~\eq{saddle} yields for these blocks:
\eqn{
  Q^{\R,\A}_\omega = - \frac 1 {\pi \nu} \sum_{\mbf k} 
  \frac 1 {\mc E (\omega) - k^2 \pm i \epsilon''(\omega)\, \omega^2 +  \frac
  {\wtilde \hbar} {2  \tau} Q^{\R,\A}_\omega}.
\label{saddle_RA}
}
The sum over the modes can be converted into an integral over $\nu\, d \mc E$, where $\mc E = k^2$. In the limit $\omega \tau \gg 1$ and \mbox{$\epsilon'' \ll \epsilon'$}, the lower integration limit can be extended to~$-\infty$. Then
$(Q^{\R,\A}_0)_\omega = \pm i$ is the solution. The full matrix can be written in the form
\eqn{
  \hat Q_0 = i \hat \Lambda, \quad
  \hat \Lambda = \l(\begin{array}{cc} 1^\R & 2F \\ 0 & -1^\A \end{array} \r),
\label{Lambda}
}
which includes the regularization in $1^{\R,\A}_\omega = e^{\pm i \omega 0^+}$ and the Keldysh block. The regularization leads to an important property $\Tr \hat Q_0^2 = 0$. 

We note that the saddle point~$\hat Q_0$ lies outside of the manifold of Hermitian matrices~$\hat Q$. The diagonal part of~$\hat Q_0$ is anti-Hermitian; this property can be traced back to the negative sign in the exponent in Eq.~\eq{e_idS}. The $\hat Q$~manifold can be continuously deformed to pass through the saddle point by making the transformation $\hat Q \mapsto e^{i \phi} \hat Q$ in the neighborhood of $\hat Q = \hat \Lambda$. As $\phi$ changes from $0$ to~$\pi/2$, the logarithm argument in Eq.~\eq{SQ} has no zero eigenvalues if $\epsilon'' > 0$. Thus, no singularities are crossed by $\exp (i S[\hat Q])$ during the deformation.

\subsection{Effective action}

The main contribution to $\langle Z \rangle$ arises from the fluctuations about the saddle point that satisfy the condition $\Tr \hat Q^2 = 0$. Such fluctuations produce weak variations of the action $S[\hat Q]$~\eq{SQ}. The matrices $\hat Q$ having the above property can be represented in the general form
\eqn{
  \hat Q (\mbf r) = i \hat R^{-1} (\mbf r)\, \hat \Lambda \hat R (\mbf r),
\label{Q_R}
}
where $\hat R$ is diagonal in the $\mbf r$~representation. 

In what follows we present the results of the calculation and refer the reader to the Appendix for details. After substituting the parameterization~\eq{Q_R} in Eq.~\eq{SQ} and omitting the $\hat Q$-independent contribution, we arrive~at
\aln{
  i S[\hat Q] &= -\Tr \ln \l(\hat 1 + \hat {\mc G} \hat \gamma \hat R 
  [\hat \gamma \hat G_0^{-1}, \hat R^{-1}] \r) \notag \\
  &\approx - \Tr\! \l(\hat {\mc G} \hat \gamma \hat R [\hat \gamma 
  \hat G_0^{-1}, \hat R^{-1}] \r) \notag \\
  &+ \frac 1 2 \Tr\! \l(\hat {\mc G} \hat \gamma \hat
  R [\hat \gamma  \hat G_0^{-1}, \hat R^{-1}] \r)^2.
\label{SQ_exp}
}
where 
\eqn{
  \hat {\mc G} = \l(\hat G_0^{-1} + i \frac {\wtilde \hbar} {2 \tau} 
  \hat \gamma \hat \Lambda \r)^{-1}
\label{G0dis}
}
is the disorder-dependent Green function operator [see Sec.~\ref{sec:corr}]. The action is expanded in the fluctuations about the saddle point, which are described by the commutator $[\hat \gamma \hat G_0^{-1}, \hat R^{-1}]$; at the saddle point $\hat R = \hat 1$ the commutator vanishes. The disorder-free inverse Green function consists of the conservative and nonconservative parts:
\eqn{
  \hat \gamma\, (\hat G_0^{-1})_\omega (\mbf k) = [\mc E (\omega) - k^2] \hat 1
  + i \epsilon''(\omega)\, \omega^2 \hat \Lambda.
\label{G0}
}
The latter results in a nontrivial contribution to the commutator due to the Keldysh structure of~$\hat \Lambda$. 

There are three leading-order contributions to~$S[\hat Q]$. Using the $\mc E (\omega)$ part of $\hat \gamma \hat G_0^{-1}$ in the linear term in Eq.~\eq{SQ_exp} we arrive~at
\eqn{
  i S_1 [\hat Q] \simeq i \pi \nu {\wtilde \hbar}\, \Tr (\pd_t \hat Q),
  \quad (\pd_t \hat Q)_{tt} \equiv (\pd_t \hat Q_{tt'} )_{t'=t}.
\label{S1}
}
The contribution of the $k^2$ part of $\hat \gamma \hat G_0^{-1}$ to the linear term of Eq.~\eq{SQ_exp} is neglected compared to its contribution to the second-order term, which gives
\eqn{
  i S_2 [\hat Q] \simeq - \frac \pi 4 \nu {\wtilde \hbar} \bar D \,
  \Tr (\pd_{\mbf r} \hat Q)^2.
\label{S2}
}
To derive this result, we used the property~\cite{akke07}
\eqn{
  \frac {\wtilde \hbar} {2 \pi \nu} \sum_{\mbf k} 
  \mc G^\R_{\omega_0} (\mbf k)\, \mc G^\A_{\omega_0} (\mbf k) \simeq 
  \l(\frac 1 \tau + \frac {2 \epsilon'' \omega_0^2} {\wtilde \hbar} \r)^{-1}
  \equiv \bar \tau 
\label{tau}
}
yielding the effective scattering time~$\bar \tau$ and defined the effective diffusion coefficient in two dimensions,
\eqn{
  \bar D = \frac 1 2 v^2 \bar \tau = \frac {2 \epsilon' \omega_0^2} 
  {\wtilde \hbar^2} \bar \tau,
\label{D}
}
where $v$~is the group velocity of light in the medium. In Sec.~\ref{sec:ld} we show that $\bar \tau$ and $\bar D$ are the relevant parameters to describe the diffusive light propagation [see Eq.~\eq{D_tau}]. Finally, the nonconservative part of $\hat \gamma \hat G_0^{-1}$ yields, in the linear order in the commutator,
\eqn{
  i S_3 [\hat Q] = \pi \nu \epsilon'' \omega_0^2\, \Tr (i \hat \Lambda \hat
  Q  + \hat \Lambda^2).
\label{S3}
}

The contributions $S_{1,2,3} [\hat Q]$ sum up to yield the NLSM effective action
\aln{
  &i S [\hat Q] = \notag \\
  &- \pi \nu_0 \Tr\! \l[-i \pd_t \hat Q + \frac {\bar D} 4 
  (\pd_{\mbf r} \hat Q)^2 - \frac {\epsilon'' \omega_0^2} {\wtilde \hbar} 
  (i \hat \Lambda \hat Q  + \hat \Lambda^2) \r].
\label{Seff}
}
The action vanishes at the saddle point, $S [i \hat \Lambda] = 0$. The key assumption behind the NLSM is the smallness of the action for fluctuations of~$\hat Q$ restricted to the manifold $\Tr \hat Q^2 = 0$, compared to the action for arbitrary fluctuations about the saddle point. The terms $S_{1,2} [\hat Q]$, which also appear in the NLSM for disordered fermionic systems~\cite{kame09}, depend only on the derivatives of~$\hat Q$. Therefore, the dominant contribution to the partition function comes from the fluctuations $\hat Q_{tt'} (\mbf r)$ [or $\hat R_{tt'} (\mbf r)$] that are slowly varying functions of $\mbf r$ and $(t + t')/2$. These ``massless modes'' are associated with the diffusive light propagation. The assumption of slow variation justifies neglecting of higher-order terms in the expansion~\eq{SQ_exp}. The contribution~$S_3 [\hat Q]$ results from the nonconservative nature of the medium. It is, in general, comparable to the to the ``massive'' $\Tr \hat Q^2$ term, unless the rate of absorption or gain is smaller than the scattering rate: 
\eqn{
  \frac {|\epsilon''| \omega_0^2} {\wtilde \hbar} \ll \frac 1 \tau.
\label{dif_reg}
}
This condition specifies the regime when the light propagation is diffusive. If this requirement is not fulfilled, the massive fluctuations beyond the NLSM have to be taken into account.

\section{Light diffusion
\label{sec:ld}}

In this section we calculate the disorder-averaged Green-function correlator. In particular, we consider the contribution that arises from the fluctuations of the field~$\hat Q$ in the neighborhood of the saddle point. The correlator possesses a diffusion-pole structure modified by a finite absorption/gain rate.

\subsection{Fluctuations about the saddle point}

We consider the parameterization 
\eqn{
  \hat Q = i \hat U e^{-\hat W/2} \hat \sigma_z e^{\hat W/2} \hat U^{-1},
  \quad \hat U = \hat U^{-1} = \l( \begin{array}{cc} 1 & F \\ 0 & -1 
  \end{array} \r),
\label{Q_W}
}
where $\hat \sigma_z$ is the Pauli matrix. Because $\hat \Lambda =  \hat U \hat \sigma_z \hat U^{-1}$ (if the regularization of unit operators is neglected), this parameterization is equivalent to Eq.~\eq{Q_R} with $\hat R = \hat U \exp(\hat W/2)\, \hat U^{-1}$. It can be verified by explicit calculation that the diagonal blocks of~$\hat W$ do not contribute to~$S [\hat Q]$ and the Green-function correlator, at least, up to the second order in~$\hat W$. We, therefore, represent this field in the form
\eqn{
  \hat W = \l( \begin{array}{cc} 0 & w \\ w^\dag & 0 \end{array} \r).
}
The specific choice of~$\hat W$ as a Hermitian matrix is justified by the requirement of convergence of the functional integral for the partition function (see below). The operator $w$ is diagonal in the $\mbf r$~representation. 

By expanding the parameterization~\eq{Q_W} in the powers of~$\hat W$ we find the first- and second-order deviations from the saddle point,
\aln{
  &\delta \hat Q^{(1)} = i \l( \begin{array}{cc} -F w^\dag & 
  - w - F w^\dag F \\ w^\dag & w^\dag F \end{array} \r), \\
  &\delta \hat Q^{(2)} = \frac i 2 \l( \begin{array}{cc} ww^\dag &
  ww^\dag F + F ww^\dag  \\ 0 & - ww^\dag \end{array} \r).
}
We note that only the latter matrix has the causality structure; however, the fluctuations of~$\hat Q$ are not required to obey causality. By using $\delta \hat Q^{(1,2)}$ in Eq.~\eq{Seff} we can calculate fluctuations of the effective action. 

The first-order variation of $S [\hat Q]$ depends on the derivatives~\footnote{To transfer the differentiation from $w$ and~$w^\dag$ to~$F$, integration by parts can be used.} of the distribution function~$F$ generated by the first two terms in Eq.~\eq{Seff}; the third term yields an identically vanishing first-order contribution. The saddle-point equation~\eq{saddle_RA} determines the retarded and advanced sectors of~$\hat Q$, but not the function~$F$. By setting to zero the variation of the effective action near the saddle point, we obtain the Usadel equation
\eqn{
  (-\pd_{\bar t} + \bar D \pd_{\mbf r}^2)\, F_{\omega_0} (\mbf r, \bar t) = 0
}
for~$F_{tt'} (\mbf r)$ in the mixed representation of the slow time variable $\bar t = (t + t')/2$ and the large frequency $\omega$, conjugate to $t-t'$. 

The second-order variation~is
\aln{
  &i \delta S^{(2)} [w, w^\dag] = - \pi \nu_0 \Tr\! \l[-i \pd_t \, \delta \hat
  Q^{(2)} + \frac {\bar D} 4  (\pd_{\mbf r} \, \delta \hat Q^{(1)})^2 \r.
  \notag \\
  &\l.+ i \frac {\bar D} 2 (\pd_{\mbf r} \hat \Lambda) (\pd_{\mbf r}\, \delta
  \hat Q^{(2)}) - i \frac {\epsilon'' \omega_0^2} {\wtilde \hbar} \hat
  \Lambda \delta \hat Q^{(2)} \r] \notag \\
  &= - \frac {\pi \nu_0} 2\! \sum_{\omega \omega' \mbf k} |w_{\omega \omega'} 
  (\mbf k)|^2\! \l[- i (\omega - \omega') + \bar D k^2 + \frac {2 \epsilon''
  \omega_0^2} {\wtilde \hbar} \r]\! ,
\label{dS2}
}
where $w (\mbf k)$ is the Fourier transform of~$w (\mbf r)$. Of the two terms with spatial gradients, the second term has a zero trace. The first term yields the $\bar D k^2$ contribution to $\delta S^{(2)}$, as well as the additional correction
\eqn{
  i \delta S^{(2)}_F [w^\dag] = - \frac {\pi \nu_0} 2 \bar D \, \text{tr}
  [w^\dag (\pd_{\mbf r} F)]^2,
}
where ``tr'' denotes the trace of operators that do not have the Keldysh matrix structure. This correction vanishes when $F(\mbf r)$ is uniform, which we will assume. With the help of Eq.~\eq{dS2}, the disorder-averaged partition function can be approximated by the functional integral
\eqn{
  \langle Z \rangle \approx \mc N_w \int D [w, w^\dag]\, 
  e^{i \delta S^{(2)}[w, w^\dag]},
}
where $\mc N_w$ is a normalization constant. For a medium with gain, the divergence of the integral for the modes with
\eqn{
  k < k_{\min} \equiv \sqrt{-\frac {2 \epsilon'' \omega_0^2} {\wtilde \hbar
  \bar D}}
\label{kmin}
}
indicates that the long-scale fluctuations become unstable due to onset of lasing (see Sec.~\ref{sec:disc}). Thus, in the long-wavelength limit the linear-gain theory breaks down and the nonlinear effects have to be taken into account~\cite{fran09,fran09b}.

\subsection{Disorder-averaged correlator
\label{sec:corr}}

Green functions and their combinations can be expressed in terms of derivatives of the partition function with respect to the source fields:
\aln{
  &G^{jk} (1,2) = - \frac i {16 \pi} \l. \frac {\delta^2 Z [\hat J,  \hat
  J^\dag]} {\delta [J^j (1)]^*\, \delta J^k (2)} \r|_{\hat J = \hat J^\dag =
  0}, \\
  &G^{jk} (1,2)\, G^{lm} (3,4) + G^{jm} (1,4)\, G^{lk} (3,2) \notag \\
  &= - \frac 1 {(16 \pi)^2} \l. \frac {\delta^4 Z [\hat J, \hat J^\dag]}
  {\delta [J^j (1)]^*\, \delta J^k (2)\, \delta [J^l (3)]^*\, \delta J^m (4)}
  \r|_{\hat J = \hat J^\dag = 0}, \\
  &Z [\hat J, \hat J^\dag] = \mc N \int\! D[\hat A, \hat A^\dag]\, 
  e^{i S [\hat A, \hat A^\dag] + \hat J^\dag \hat A + \hat A^\dag  \hat J},
\label{ZJ}
}
where $j,k,l,m = \cl,\q$ and $1,2, \ldots$ are full sets of coordinates in some representation, e.g., $1 = (\mbf k_1, \omega_1)$, etc. By inverting the matrix~\eq{G-1_matr}, we identify the sectors of the Green function as $G^{\cl, \q} = G^\R$, $G^{\q, \cl} = G^\A$, $G^{\cl, \cl} = G^\K \ne [(G^{-1})^\K]^{-1}$, and $G^{\q, \q} = 0$. 

The disorder-averaged Green functions and correlators are obtained by using the above expressions with the disorder-averaged partition function~\footnote{Equation~\eq{ZJdis} is derived analogously to Eq.~\eq{Zdis} by adding the source terms, as in Eq.~\eq{ZJ}, before integrating out the fields $\hat A$ and~$\hat A^\dag$.}
\aln{
  &\langle Z [\hat J, \hat J^\dag] \rangle = \wtilde {\mc N}_Q \!\int \!
  D [\hat Q]\, \exp\! \l(i S[\hat Q] + 16 \pi i \hat J^\dag \hat {\mc G}_{\hat
  Q}  \hat J \r)\!, 
\label{ZJdis}\\
  &\hat {\mc G}_{\hat Q} \equiv \l(\hat G_0^{-1} + \frac {\wtilde \hbar} {2
  \tau}  \hat \gamma \hat Q \r)^{-1}.
\label{GQ}
}
We find, in particular,
\aln{
  &\langle G^{\R,\A,\K} (1,2) \rangle = \langle \mc G^{\R,\A,\K}_{\hat Q} (1,2)
  \rangle_{\hat Q}, 
\label{Gdis}\\
  &\langle G^\R (1,2)\, G^\A (3,4) \rangle \notag \\
  &= \langle \mc G^\R_{\hat Q} (1,2)\, \mc G^\A_{\hat Q} (3,4) + \mc
  G^\K_{\hat Q} (1,4)\, \mc G^\Q_{\hat Q} (3,2) \rangle_{\hat Q},
\label{Gcorr}
}
where the average $\langle \cdots \rangle_{\hat Q}$ over~$\hat Q$ is performed with the exponential weight $\exp (i S[\hat Q])$. Equation~\eq{Gdis} shows that $\hat {\mc G} = \hat {\mc G}_{i \hat \Lambda}$ [Eq.~\eq{G0dis}] is the disorder-averaged Green function in the lowest-order saddle-point approximation. The component \mbox{$\mc G^\Q_{\hat Q} \equiv \mc G^{\q,\q}_{\hat Q}$} in Eq.~\eq{Gcorr} is, in general, non-zero when $\hat Q$ does not have the causality structure. This observation is essential for the following calculation. 

We calculate the Green-function correlator
\eqn{
  \mc R (1,2,3,4) \equiv \langle G^\R (1,2)\, G^\A (3,4) \rangle -  \langle
  G^\R (1,2)  \rangle \langle G^\A (3,4) \rangle
} 
by expansion about the saddle point. The lowest-order correction to the Green function~\eq{GQ}~is
\aln{
  &\hat {\mc G}_{\hat Q} - \hat {\mc G} \simeq - \frac {\wtilde \hbar} {2
  \tau} \hat {\mc G} \hat \gamma \delta \hat Q^{(1)} \hat {\mc G} \notag \\
  &= i \frac {\wtilde \hbar} {2 \tau} \l( \begin{array}{cc} 
  \mc G^\R w \mc G^\A + F \mc G^\A w^\dag \mc G^\R F & F \mc G^\A w^\dag 
  \mc G^\R \\ -\mc G^\A w^\dag \mc G^\R F & -\mc G^\A w^\dag \mc G^\R 
  \end{array} \r).
\label{GQ1}
}
The Gaussian averages with the action~\eq{dS2} are as follows:
\aln{
  &\langle w \rangle_w = \langle w^\dag \rangle_w = 0, \quad
  \langle w^\dag (1,2)\, w^\dag (3,4) \rangle_w = 0 \\
  &\langle w (1,2)\, w^\dag (3,4) \rangle_w  \notag \\
  &= \frac 2 {\pi \nu_0}  \frac {\delta_{\mbf k_1 - \mbf k_2, \mbf  k_4 - \mbf
  k_3} \delta_{\omega_1, \omega_4} \delta_{\omega_2, \omega_3}} {\bar D (\mbf
  k_1 - \mbf k_2)^2 -i (\omega_1 - \omega_2) + \frac {2 \epsilon''
  \omega_0^2} {\wtilde \hbar}}.
}
Therefore, the leading contribution to the correlator comes from the K-Q term in Eq.~\eq{Gcorr}, which is given by the product of diagonal blocks in Eq.~\eq{GQ1}. We find
\aln{
  &\mc R (1,2,3,4) = \frac {\wtilde \hbar} {2 \pi \nu \tau^2} 
  \mc G^\R (1)\, \mc G^\R (2)\, \mc G^\A (3)\, \mc G^\A (4) \notag \\
  &\times \frac {\delta_{\mbf k_1 - \mbf k_4, \mbf  k_2 - \mbf k_3} 
  \delta_{\omega_1, \omega_2} \delta_{\omega_4, \omega_3}} 
  {\bar D (\mbf k_1 - \mbf k_4)^2 -i (\omega_1 - \omega_4) + \frac {2
  \epsilon'' \omega_0^2} {\wtilde \hbar}}.
}
The correlator has a diffusion pole with the diffusion coefficient~$\bar D$. The pole is modified by the $\epsilon''$~term that arises from the corresponding contribution in the effective action~\eq{Seff}. This term defines the absorption rate
\eqn{
  \frac 1 {\tau_{\text a}} \equiv \frac {2 \epsilon'' \omega_0^2} {\wtilde
  \hbar},
}
negative for gain.

\subsection{Discussion
\label{sec:disc}}

The pole structure of the correlator implies that the light intensity~$I$ in the medium satisfies the diffusion equation with a nonconservative term:
\aln{
  &\l(\pd_t - \bar D \nabla^2 + \tau_{\text a}^{-1} \r) I = 0, \\
  &\bar D = \frac 1 2 v^2 \bar \tau = \frac 1 2 v^2 \l(\frac 1 \tau + 
  \frac 1 {\tau_{\text a}}\r)^{-1}.
}
We compare this equation with
\aln{
  &\l[\frac {\tau} {1 + 2 \tau/ \tau_{\text a}} \pd_t^2 + \pd_t - D' \nabla^2 
  + \tau_{\text a}^{-1} \frac {1 + \tau/ \tau_{\text a}} 
  {1 + 2 \tau/ \tau_{\text a}} \r] I = 0, 
\label{difeq_tr}\\
  &D' \equiv \frac 1 2 v^2 \l(\frac 1 \tau + \frac 2 {\tau_{\text a}}\r)^{-1},
}
that follows from the photon transport equation (see Eq.~(15) of Ref.~\cite{durd97}). According to Ref.~\cite{durd97}, the light propagation is diffusive if the second derivative with respect to time in Eq.~\eq{difeq_tr} can be neglected. This is the case when
\eqn{
  \tau \ll \Delta t,
\label{dif_cond}
}
where $\Delta t$ is the characteristic time scale of intensity variation. The reaction of the medium on a fluctuation of intensity will be determined by the shortest time scale, so that $\Delta t \lesssim |\tau_{\text a}|$ can be assumed. Therefore, when neglecting the corrections of the order of $\tau / \Delta t$ in Eq.~\eq{difeq_tr}, we also have to neglect the contributions of the order of~$\tau / \tau_{\text a}$. In particular, it is consistent with the diffusion approximation to set 
\eqn{
  D' \simeq D \equiv \frac 1 2 v^2 \tau. 
}
The independence of absorption for the diffusion coefficient was also supported by the numerical evidence in Ref.~\cite{durd97}. It is worth commenting on the claim~\cite{furu94,furu97} that the diffusion coefficient in the medium with absorption must be equal to~$D$ even for $\tau / \tau_{\text a} \sim 1$. A closer look at the derivation of the diffusion coefficient from the transport equation in Ref.~\cite{furu94} reveals that the time-derivative terms neglected in Eqs.~(A9) and~(A11) of that article would yield the diffusion coefficient 
\eqn{
  D'' = \frac 1 2 v^2 \tau \l(1 - 2 \frac \tau {\tau_{\text a}} \r) \simeq D' 
}
were they taken into account. Thus, the (approximate) independence of the diffusion coefficient of absorption is a consequence of the self-consistent application of the diffusion-approximation conditions~\eq{dif_reg} and~\eq{dif_cond}.

The NLSM effective action~\eq{Seff} is derived under the condition~\eq{dif_cond} as well. This condition guarantees the slow variation of~$\hat Q$, and makes it possible to neglect the contribution of $\mc E (\omega)$ part of $\hat \gamma \hat G_0^{-1}$ [Eq.~\eq{G0}] to the second-order term in Eq.~\eq{SQ_exp}. This contribution would result in a second-time-derivative term in the effective action. Again, the diffusion approximation requires that we set
\eqn{
  \bar D \simeq D, \quad \bar \tau \simeq \tau
\label{D_tau}
}
in the NLSM expressions. Thus, the NLSM and the theory of transport equation agree in the diffusive regime.

In the medium with gain, the diffusive relaxation competes with the amplification. Because the long-scale intensity fluctuations disperse slower, they become unstable, and the random lasing sets in. The cutoff wavenumber $k_{\min}$~\eq{kmin} determines the critical sample size 
\eqn{
  l = \sqrt{D\,  |\tau_{\text a}|}
}
above which the system is lasing and the linear-gain theory does not apply. Alternatively, the above expression yields the lasing-threshold value of $|\tau_{\text a}|$ if $l$ is given.

\section{Conclusions}

We obtained the functional-integral form of the partition function for an optical medium with linear absorption or gain. Keldysh technique is particularly suitable for description of nonconservative systems because it provides a natural representation for the action. The \emph{disorder-averaged} partition function is expressed as a functional integral over the auxiliary matrix field~$\hat Q$. Within the framework of nonlinear sigma model, we considered the fluctuations about the saddle point that fulfill the condition $\Tr \hat Q^2 = 0$. We found that the effective action $S[\hat Q]$ for these fluctuations contains an extra term due to absorption or gain. 

With the help of the nonlinear-sigma-model partition function, we computed the disorder-averaged Green-function correlator. The leading contribution from the vicinity of the saddle point has the diffusion-pole structure modified by a finite absorption/gain rate. The diffusion coefficient is found to be approximately independent of the absorption or gain in agreement with the theory of photon transport equation. In the medium with gain, the linear theory is not applicable in the long-wavelength limit. If the sample size exceeds a certain critical length, the random lasing sets in.

\appendix*

\section{Derivation of the effective action}

\subsection{Derivation of Eq.~\eq{SQ_exp}}

After substitution of the parameterization~\eq{Q_R} in Eq.~\eq{SQ} we obtain
\aln{
  i S[\hat Q] &= -\Tr \ln \l[\hat R \hat \gamma \hat R^{-1} \l( \hat R 
  \hat \gamma \hat G_0^{-1} \hat R^{-1} + i \frac {\wtilde \hbar} {2 \tau} 
  \hat \Lambda \r) \r] \notag \\
  &= -\Tr \ln \hat \gamma - \Tr \ln \l(\hat \gamma \hat {\mc G}^{-1} +
  \hat R [\hat \gamma \hat G_0^{-1}, \hat R^{-1}] \r).
}
By separating $\Tr \ln (\hat \gamma \hat {\mc G}^{-1})$ and dropping the $\hat Q$-independent terms we arrive at Eq.~\eq{SQ_exp}.

\subsection{Derivation of Eq.~\eq{S1}}

The conservative part of $\hat \gamma \hat G_0^{-1}$, when substituted in the first trace in Eq.~\eq{SQ_exp}, yields
\aln{
  i S_1 [\hat Q] = &-\Tr\! \sum_{\substack{\omega, \omega'\\ \mbf k,\mbf k'}} 
  \hat {\mc G}_\omega (\mbf k) \hat \gamma \hat R_{\omega \omega'} (\mbf k -
  \mbf k') \, \hat R^{-1}_{\omega' \omega} (\mbf k' - \mbf k)  \notag \\
  &\times [\mc E (\omega') - \mc E (\omega) - k'^2 + k^2] \notag \\
  \simeq - \Tr\! \sum_{\substack{\omega, \Delta \omega\\ \mbf k, 
  \Delta \mbf k}}
  &\hat {\mc G}_\omega (\mbf k) \hat \gamma \hat R_{\omega, 
  \omega + \Delta \omega} (- \Delta \mbf k)\, \hat R^{-1}_{\omega + 
  \Delta \omega, \omega} (\Delta \mbf k)\  \notag \\
  &\times [\wtilde \hbar \Delta \omega  - (2 \mbf k + \Delta \mbf k) 
  \cdot \Delta \mbf k],
}
where $\Delta \omega = \omega' - \omega$ and $\Delta \mbf k = \mbf k' - \mbf k$. We note that $\hat R$ is peaked at small wave vectors in the $k$~representation. The sum 
\eqn{
  \sum_{\mbf k} \hat {\mc G}_\omega (\mbf k) \hat \gamma = - i \pi \nu 
  \hat \Lambda
\label{sumG}
} 
follows from the saddle-point condition; furthermore, $\sum_{\mbf k} \hat {\mc G}_\omega (\mbf k) \, \mbf k = 0$ due to the symmetry. After calculating the $\Delta \mbf k$ sum we arrive~at
\aln{
  &i S_1 [\hat Q] = i \pi \nu\, \Tr\! \sum_{\omega, \Delta \omega} \int d \mbf
  r \hat \Lambda \hat R_{\omega,  \omega + \Delta \omega} (\mbf r)  \notag \\
  &\times [\wtilde \hbar \Delta \omega + \pd_{\mbf r}^2]\,
  \hat R^{-1}_{\omega + \Delta \omega, \omega} (\mbf r)  \notag \\
  &= \pi \nu\, \Tr\! \int d \mbf r\, dt\, dt'\, i \hat \Lambda \hat R_{t't}
  (\mbf r)  [i \wtilde \hbar \pd_t + \pd_{\mbf r}^2]\, \hat R^{-1}_{tt'}  (\mbf
  r).
}
Applying the representation~\eq{Q_R} we obtain Eq.~\eq{S1} from the $\pd_t$ part. The $\pd_{\mbf r}^2$ part is neglected compared to~$i S_2 [\hat Q]$; the latter contribution is multiplied by ${\wtilde \hbar} \bar D \sim \omega_0 \bar \tau \gg 1$.

\subsection{Derivation of Eq.~\eq{S2}}

We substitute the $k^2$ part of $\hat \gamma \hat G_0^{-1}$ in the second trace in Eq.~\eq{SQ_exp} to get 
\aln{
  &i S_2 [\hat Q] = \frac 1 2 \Tr\!\! \sum_{\mbf k_1 \ldots \mbf k_4}  \hat
  {\mc G} (\mbf k_1) \hat \gamma\, \hat R (\mbf k_1 - \mbf k_2)\, \hat R^{-1}
  (\mbf k_2 - \mbf k_3) \notag \\
  &\times  
  \hat {\mc G} (\mbf k_3) \hat \gamma\, \hat R
  (\mbf k_3 - \mbf k_4)\, \hat R^{-1} (\mbf k_4 - \mbf k_1) 
  (k_2^2 - k_3^2) (k_4^2 - k_1^2) \notag \\
  &\simeq 2 \Tr\!\! \sum_{\substack{\bar {\mbf k}\, \Delta \bar {\mbf k}\\ 
  \Delta \mbf k\, \Delta \mbf k'}} \hat {\mc G} (\bar {\mbf k}) \hat \gamma\,
  \hat R \l(\Delta \bar {\mbf k} - \frac {\Delta \mbf k + \Delta \mbf k'} 2 \r)
  \hat R^{-1} (\Delta \mbf k') \notag \\
  &\times \hat {\mc G} (\bar {\mbf k}) \hat \gamma\, \hat R \l(-\Delta 
  \bar {\mbf k} - \frac {\Delta \mbf k + \Delta \mbf k'} 2 \r) 
  \hat R^{-1} (\Delta \mbf k) \notag \\
  &\times (\bar {\mbf k} \cdot \Delta \mbf k) (\bar {\mbf k} \cdot 
  \Delta \mbf k'),
}
where $\bar {\mbf k} = \sum_{i=1}^4 \mbf k_i/4$ and we neglected the contributions of higher order in $\Delta \bar {\mbf k} = (\mbf k_1 + \mbf k_4 - \mbf k_2 - \mbf k_3)/2$, $\Delta \mbf k = \mbf k_4 - \mbf k_1$, and~$\Delta \mbf k' = \mbf k_2 - \mbf k_3$. We use the representation
\eqn{
  \hat {\mc G} \hat \gamma = \frac 1 2 \mc G^\R \, (\hat 1 + \hat \Lambda)
  + \frac 1 2 (\hat 1 - \hat \Lambda)\, \mc G^\A
}
and the well-known relations [see Eqs.~\eq{D} and~\eq{tau}]
\aln{
  &\sum_{\mbf k} \mc G^\R_{\omega} (\mbf k)\, \mc G^\A_{\omega'} (\mbf k)\,
  k_\alpha k_\beta \simeq \frac 1 2 \pi \nu \wtilde \hbar \bar D 
  \delta_{\alpha \beta}, \\
  &\sum_{\mbf k} \mc G^{\R (\A)}_{\omega} (\mbf k)\, \mc G^{\R (\A)}_{\omega'}
  (\mbf k)\, k_\alpha k_\beta \simeq 0,
}
to find 
\aln{
  &i S_2 [\hat Q] = - \frac 1 2 \pi \nu \wtilde \hbar \bar D \, \Tr [ (\hat 1 +
  \hat \Lambda) \hat R\, (\pd_{\mbf r}  \hat R^{-1}) \notag \\
  &\cdot (\hat 1 - \hat \Lambda) \hat R \, (\pd_{\mbf r} \hat R^{-1})]
  = \frac 1 4 \pi \nu \wtilde \hbar \bar D \, \Tr [\pd_{\mbf r} (\hat R^{-1}
  \hat \Lambda \hat R)]^2,
}
from which Eq.~\eq{S2} follows.

\subsection{Derivation of Eq.~\eq{S3}}

The nonconservative part of $\hat \gamma \hat G_0^{-1}$, being substituted in the first trace in Eq.~\eq{SQ_exp} yields
\aln{
  i S_3 [\hat Q] \simeq &- i \epsilon'' \omega_0^2 \, \Tr \! \sum_{\mbf k
  \mbf k'} \hat {\mc G} (\mbf k) \hat \gamma \hat R (\mbf k - \mbf k') 
  \notag \\
  &\times [ \hat \Lambda \hat R^{-1} (\mbf k' - \mbf k) - \hat R^{-1} (\mbf k'
  - \mbf k)\, \hat \Lambda].
}
We change the variable $\mbf k' = \mbf k + \Delta \mbf k$ and apply Eq.~\eq{sumG}. After cyclically moving the operators under the trace we obtain Eq.~\eq{S3}.

% If you have acknowledgments, this puts in the proper section head.
\begin{acknowledgments}
This work was supported by a grant from the Kyung Hee University in~2011 (grant No. KHU-20110683). We thank Johann Kroha for helpful discussions. 

\end{acknowledgments}

\bibliography{transport.bib}

\end{document}